\documentclass[12pt]{article}

\usepackage{amsmath, amssymb, slashed, epsf, color, subfigure, ulem}

\usepackage{graphicx}

\usepackage{color}

\makeatletter
\renewcommand{\section}{%
  \@startsection{section}%
   {1}%
   {\z@}%
   {-2.5ex \@plus -1ex \@minus -.2ex}%
   {2.3ex \@plus.2ex}%
   {\normalfont\large\bfseries}%
}%
\makeatother

\makeatletter
\renewcommand{\subsection}{%
  \@startsection{subsection}%
   {1}%
   {\z@}%
   {-2.5ex \@plus -1ex \@minus -.2ex}%
   {2.3ex \@plus.2ex}%
   {\normalfont\normalsize\bfseries}%
}%
\makeatother

\begin{document}


\def\a{\alpha}
\def\b{\beta}
\def\c{\varepsilon}
\def\d{\delta}
\def\e{\epsilon}
\def\f{\phi}
\def\g{\gamma}
\def\h{\theta}
\def\k{\kappa}
\def\l{\lambda}
\def\m{\mu}
\def\n{\nu}
\def\p{\psi}
\def\q{\partial}
\def\r{\rho}
\def\s{\sigma}
\def\t{\tau}
\def\u{\upsilon}
\def\v{\varphi}
\def\w{\omega}
\def\x{\xi}
\def\y{\eta}
\def\z{\zeta}
\def\D{\Delta}
\def\G{\Gamma}
\def\H{\Theta}
\def\L{\Lambda}
\def\F{\Phi}
\def\P{\Psi}
\def\S{\Sigma}

\def\o{\over}
\newcommand{\gsim}{ \mathop{}_{\textstyle \sim}^{\textstyle >} }
\newcommand{\lsim}{ \mathop{}_{\textstyle \sim}^{\textstyle <} }
\newcommand{\vev}[1]{ \left\langle {#1} \right\rangle }
\newcommand{\bra}[1]{ \langle {#1} | }
\newcommand{\ket}[1]{ | {#1} \rangle }
\newcommand{\EV}{ {\rm eV} }
\newcommand{\KEV}{ {\rm keV} }
\newcommand{\MEV}{ {\rm MeV} }
\newcommand{\GEV}{ {\rm GeV} }
\newcommand{\TEV}{ {\rm TeV} }
\newcommand{\1}{\mbox{1}\hspace{-0.25em}\mbox{l}}
\def\diag{\mathop{\rm diag}\nolimits}
\def\Spin{\mathop{\rm Spin}}
\def\SO{\mathop{\rm SO}}
\def\O{\mathop{\rm O}}
\def\SU{\mathop{\rm SU}}
\def\U{\mathop{\rm U}}
\def\Sp{\mathop{\rm Sp}}
\def\SL{\mathop{\rm SL}}
\def\tr{\mathop{\rm tr}}

\def\IJMP{Int.~J.~Mod.~Phys. }
\def\MPL{Mod.~Phys.~Lett. }
\def\NP{Nucl.~Phys. }
\def\PL{Phys.~Lett. }
\def\PR{Phys.~Rev. }
\def\PRL{Phys.~Rev.~Lett. }
\def\PTP{Prog.~Theor.~Phys. }
\def\ZP{Z.~Phys. }

\def\dd{\mathrm{d}}
\def\ff{\mathrm{f}}
\def\BH{{\rm BH}}
\def\inf{{\rm inf}}
\def\ev{{\rm evap}}
\def\eq{{\rm eq}}
\def\SM{{\rm sm}}
\def\Mpl{M_{\rm Pl}}
\def\GeV{{\rm GeV}}
\newcommand{\Red}[1]{\textcolor{red}{#1}}

\def\mDM{m_{\rm DM}}
\def\mphi{m_{\phi}}
\def\TeV{{\rm TeV}}
\def\Gphi{\Gamma_\phi}
\def\TR{T_{\rm RH}}
\def\Br{{\rm Br}}
\def\DM{{\rm DM}}
\def\Eth{E_{\rm th}}
\newcommand{\lmk}{\left(}  
\newcommand{\rmk}{\right)}
\newcommand{\lkk}{\left[}  
\newcommand{\rkk}{\right]}
\newcommand{\lhk}{\left \{ }  
\newcommand{\rhk}{\right \} }
\newcommand{\del}{\partial}  
\newcommand{\la}{\left\langle} 
\newcommand{\ra}{\right\rangle}

\def\slashchar#1{\setbox0=\hbox{$#1$} 
\dimen0=\wd0 
\setbox1=\hbox{/} \dimen1=\wd1 
\ifdim\dimen0>\dimen1 
\rlap{\hbox to \dimen0{\hfil/\hfil}} 
#1 
\else 
\rlap{\hbox to \dimen1{\hfil$#1$\hfil}} 
/ 
\fi}


\begin{titlepage}
\begin{center}

\hfill CTPU-15-30\\

\vspace{1.0cm}
{\large\bf
Diphoton channel at the LHC experiments\\ to find a hint for a new heavy gauge boson
}

\vspace{1.0cm}
{\bf Kunio Kaneta} $^a$,
{\bf Subeom Kang} $^{a,b}$,
{\bf Hye-Sung Lee} $^a$

\vspace{1.0cm}
{\it
$^a${Center for Theoretical Physics of the Universe, IBS, Daejeon 34051, Korea} \\
$^b${Department of Physics, KAIST, Daejeon 34141, Korea}
}

\vspace{1.0cm}
\abstract{
Recently there has been a huge interest in the diphoton excess around 750 GeV reported by both ATLAS and CMS collaborations,
although the newest analysis with more statistics does not seem to support the excess.
Nevertheless, the diphoton channel at the LHC experiments are a powerful tool to probe a new physics.
One of the most natural explanations of a diphoton excess, if it occurs, could be a new
scalar boson with exotic colored particles.
In this setup, it would be legitimate to ask what is the role of this new scalar in nature.
A heavy neutral gauge boson ($Z'$) is one of the traditional targets of
the discovery at the collider experiments with numerous motivations.
While the Landau-Yang theorem dictates the diphoton excess
cannot be this spin-1 gauge boson, there is a strong correlation of a
new heavy gauge boson and a new scalar boson which provides a mass to
the gauge boson being at the same mass scale.
In this paper, we point out a simple fact that a new scalar with a property similar to the recently highlighted 750 GeV
would suggest an existence of a TeV scale $Z'$ gauge boson that might be
within the reach of the LHC Run 2 experiments. We take a scenario of
the well-motivated and popular gauged $B-L$ symmetry and require the
gauge coupling unification to predict the mass and other properties of
the $Z'$ and illustrate the discovery of the $Z'$ would occur during
the LHC experiments.
}

\end{center}
\end{titlepage}
\setcounter{footnote}{0}

\section{Introduction}
In late 2015, both ATLAS and CMS have reported the existence of a diphoton excess with an invariant mass around 750 GeV with signals at the level of 3.6$\s$ (ATLAS) and 2.6$\s$ (CMS)~\cite{ATLAS,CMS}.
During the Moriond 2016 conference ATLAS and CMS have reported new results that show slightly increased statistical significance of the diphoton excess \cite{Moriond2016}.
The newest analysis with more statistics released during the ICHEP 2016 conference does not support this diphoton excess \cite{ICHEP}.

Although it seems the 750 GeV diphoton excess may not be the real signal at this point, it is interesting to see the extensive studies of this diphoton excess have appeared; models of a composite state/a pseudo Numb-Goldstone boson~\cite{diphoton:composite/NGB}, extra dimensional models~\cite{diphoton:extra dimension}, models including additional charged fermions/scalars~\cite{diphoton:additional charged particles}, models with an extended gauge symmetry~\cite{diphoton:extended gauge sector}, models with an extended Higgs sector~\cite{diphoton:extended Higgs sector}, possible link with the dark matter~\cite{diphoton:dark matter} and the model independent studies~\cite{diphoton:model independent}.
They suggest the diphoton excess search at the LHC experiments is a powerful tool to look for a new physics beyond the standard model (SM).

On the other hand, there have been huge efforts in building models and experimental searches for a new heavy neutral gauge boson ($Z'$) motivated from various contexts \cite{Langacker:2008yv}.
Although the Landau-Yang theorem \cite{LandauYang} prohibits a heavy vector boson to be the source of the diphoton excess, there may be a strong link between the $Z'$ and the diphoton excess.
A new heavy gauge boson requires a new mass generation mechanism as the SM Higgs boson cannot explain its mass.
Although there might be other ways to generate the mass such as the Stueckelberg mechanism \cite{Stueckelberg:1900zz}, a new scalar boson at the same mass scale would be a natural companion of a new gauge boson in the same manner the SM Higgs boson of the weak scale is a companion of the SM weak gauge bosons.
Therefore a discovery of a new scalar boson might be a hint of the existence of a new gauge boson at the same scale, whose discovery might be just around the corner perhaps using the same high-energy collider machine.

In this paper, we discuss that a heavy scalar field $S$ whose role is to give a mass to the same scale $Z'$ can be a good candidate to explain the diphoton excess, if observed.
For the illustration purpose, we will assume a new scalar $S$ has a similar property as the aforementioned 750 GeV scalar.
We will take its mass, width, and branching ratios are consistent with the reported data of 2015 \cite{ATLAS,CMS} in our quantitative study.
We will basically assume the situation the 750 GeV diphoton excess is still valid, although in reality it does not seem to be that way anymore as we discussed in the beginning, instead of taking a new heavy scalar with somewhat similar but different property.
For definiteness, we take the $Z'$ as a gauge boson of the $\U(1)_{B-L}$ gauge symmetry although our analysis can apply to other choices of the $\U(1)$ gauge symmetry.
The $B-L$ is one of the most natural gauge extensions of the SM as it does not require any fermionic matter particles for the chiral anomaly cancellation except for the three right-handed neutrinos \cite{WeinbergII} that are highly motivated from the nonzero neutrino masses.

As is clear from the success of the electroweak theory, it is worthwhile to look for possible ways leading to the grand unified theory (GUT)~\cite{Georgi:1974sy}.
The GUT can be a guiding principle for physics beyond the standard model, in which the SM gauge interactions are unified at a very high energy scale.
As is also well known, however, the SM fails to achieve the coupling unification by simply extrapolating the running gauge couplings toward a very high energy.
Therefore, the ideas of the GUT inevitably require new charged particles with masses above the electroweak scale but below the unification scale.\footnote{General discussion on building models with the gauge coupling unification by introducing charged particles is shown in Ref.~\cite{Giudice:2012zp}.}

When we introduce the $\U(1)_{B-L}$ and require the gauge coupling unification, the implication of a new heavy scalar for the TeV scale $Z'$ property becomes highly specific and we can estimate the mass of the $Z'$ in this scenario.
As we will see, within perturbative models, a large number of colored exotics are necessary to fit the cross section of the 750 GeV diphoton excess, otherwise the $B-L$ breaking scale would be too small to be consistent with the current experimental constraints.

In this paper, we give a possible link between the $Z'$ and the possible diphoton excess, taking a recently issued 750 GeV case as an example, and show a viable model which predicts the TeV scale $Z'$.
The LHC experiments, according to our study, has a high potential to discover this TeV scale $Z'$ in its Run 2.

The outline of the rest of this paper is given as following.
In Sec.~\ref{sec:gauge coupling unification}, we discuss the gauge coupling unification briefly.
In Sec.~\ref{sec:the model}, we present our working model to illustrate our point.
In Sec.~\ref{sec:diphoton}, we estimate the 750 GeV diphoton cross section in the model.
In Sec.~\ref{sec:implications for new gauge boson}, we predict the $Z'$ mass from the assumed 750 GeV diphoton cross section and the gauge coupling unification.
In Sec.~\ref{sec:other experimental constraints}, we discuss the constraints on the other decay modes of the 750 GeV scalar.
In Sec.~\ref{sec:discussion}, we discuss the validity of our analysis and show how generic our conclusions are.
In Sec.~\ref{sec:summary}, we summarize our study.

\section{Gauge coupling unification}
\label{sec:gauge coupling unification}
We take the GUT as a guiding principle, and impose the gauge coupling unification by introducing new charged particles at TeV scale. 
In what follows, let us consider the case that the new particles have the universal mass scale $M_*$.
At the one-loop level, the SM gauge coupling constants at the GUT scale can be written as
\begin{eqnarray}
  \a^{-1}_a(M_{\rm GUT}) &=& \a^{-1}_a(M_Z) - \frac{b_a^{\rm SM}}{2\pi}\ln\frac{M_*}{M_Z}-\frac{\D_a}{2\pi}\ln\frac{M_{\rm GUT}}{M_*},~~~(a=1,2,3),
  \label{eq:alpha_U}
\end{eqnarray}
where $\a_a=g_a^2/(4\pi)$ with $g_a$'s being the three gauge coupling constants of the SM~\cite{Giudice:2012zp}.
The parameter $b_a$'s are the coefficients of the beta functions, and $b_a^{\rm SM}$ denotes their SM value: $(b_1^{\rm SM},b_2^{\rm SM},b_3^{\rm SM})=(41/10,-19/6,-7)$.
At the energy scales above $M_*$, new charged particles change $b_a$'s, which are parametrized by $\D_a\equiv b_a-b_a^{\rm SM}$ with $b_a$'s including new particle contributions.

\begin{figure}[t]
\begin{center} 
    \includegraphics[width=0.5\textwidth]{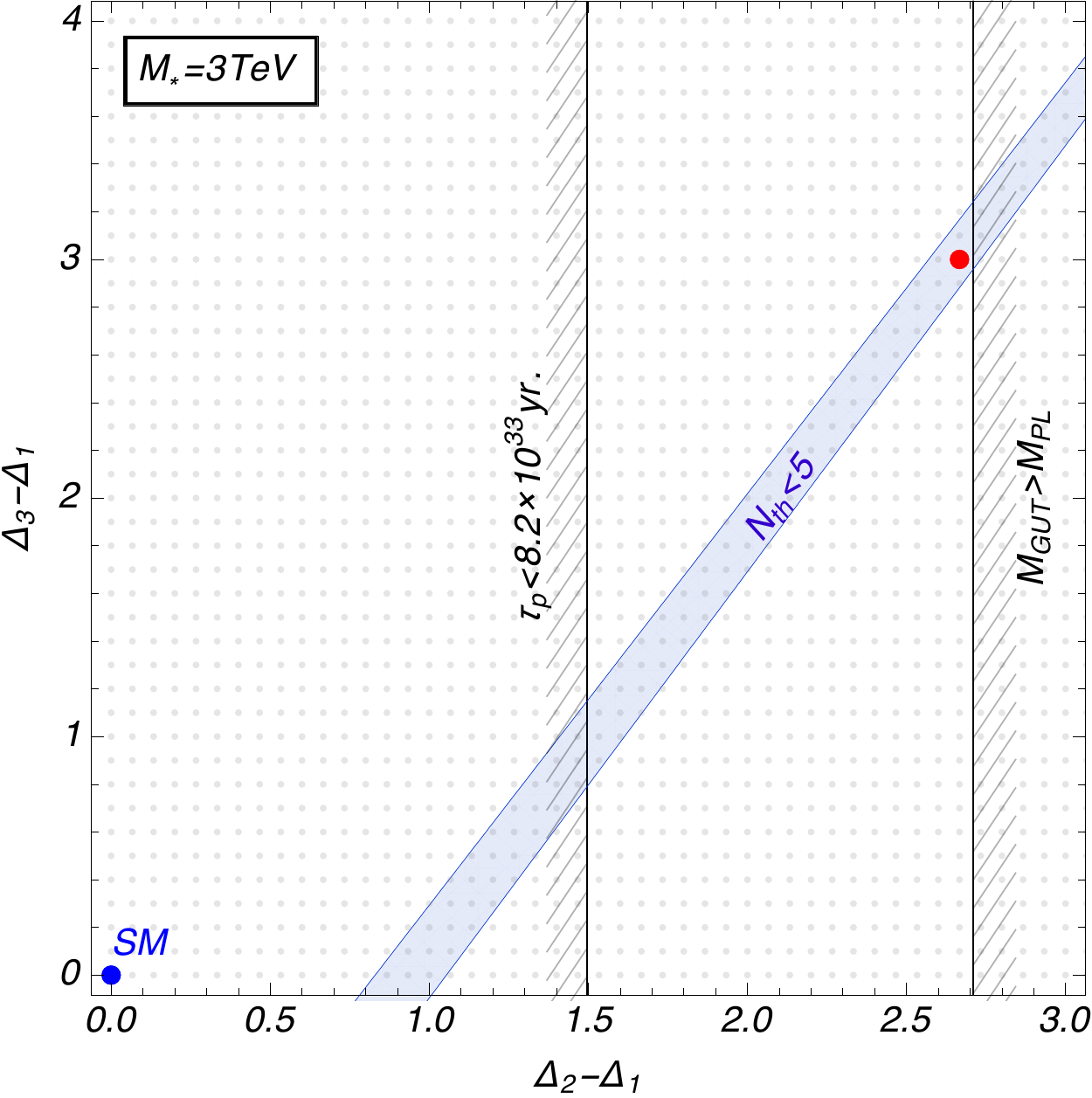}
    \caption{Each gray dot represents a possible choice of $\D_2-\D_1$ and $\D_3-\D_1$. We take $M_*=3~\TEV$. The blue region is where the condition of $N_{\rm th}<5$ (for satisfactory gauge coupling unification) holds. The regions $\D_2-\D_1\lesssim1.5$ and $\D_2-\D_1\gtrsim2.7$ are excluded by the rapid proton decay and the requirement of $M_{\rm GUT}<M_{\rm PL}$, respectively. The red point $(\D_2-\D_1,\D_3-\D_1)=(8/3,3)$ corresponds to the model considered in Sec.~\ref{sec:the model}.}
    \label{fig:GCU_map}
\end{center}
\end{figure}

From a viewpoint of the GUT idea, low energy gauge couplings are determined by a single gauge coupling at the GUT scale by taking into account appropriate threshold corrections.
We need to specify a concrete GUT model in order to evaluate the threshold corrections.
In this study, however, instead of specifying a GUT model, we show how large corrections are required to unify the gauge couplings by utilizing
\begin{eqnarray}
  N_{\rm th} &\equiv&
  2\pi|\D\a^{-1}|
  =
  2\pi|\a_{1,2}^{-1}(M_{\rm GUT})-\a_3^{-1}(M_{\rm GUT})|,
\end{eqnarray}
where we define the GUT scale $M_{\rm GUT}$ to be $\a_{1}^{-1}(M_{\rm GUT})=\a_{2}^{-1}(M_{\rm GUT})$.
For instance, the threshold parameters in the supersymmetric SM lead to $N_{\rm th}\lesssim 5$~\cite{Bagger:1995bw} when superparticles are around TeV scale.
Once we fix the universal mass scale $M_*$, the $M_{\rm GUT}$ can be determined by the value of the difference $\D_2-\D_1$ from Eq.~(\ref{eq:alpha_U}).
Then, we can evaluate $N_{\rm th}$ by taking a certain value of $\D_3-\D_{1,2}$.
Figure~\ref{fig:GCU_map} shows all possible choices of $\D_2-\D_1$ and $\D_3-\D_1$, for a case of $M_*=3~\TEV$.
(This $M_*$ value is motivated from the dark matter relic density constraint as will be explained in the following section.)
The SM corresponds to $(\D_2-\D_1,\D_3-\D_1)=(0,0)$ in the figure.
We show $N_{\rm th}<5$ region as a reference, which is depicted by the blue region.
For upper bound on $\D_2-\D_1$, we require that the GUT scale should be smaller than the Planck scale $M_{\rm PL}\simeq 1.22\times 10^{19}~\GEV$ so that the unification is realized perturbatively.
The lower bound on $\D_2-\D_1$ comes from the observed limit of the proton life time, which we take $\t_p\sim \a_U^{-2}m_p^{-5}M_{\rm GUT}^4 > 8.2\times 10^{33}$~yr.~\cite{Nishino:2012bnw}, where $\a_U\equiv \a_{1,2}(M_{\rm GUT})$ and $m_p$ is the proton mass.
Here, we have assumed that only the GUT gauge bosons with a mass of $M_{\rm GUT}$ may cause the proton decay such as $p\to\pi^0 e^+$.

\section{Model}
\label{sec:the model}
There exist a large number of possible models that can realize the unification, and also explain the diphoton excess at the LHC.
Instead of listing such possible models, we restrict ourselves to the case that the additional particles are only squark-like scalar particles ($\tilde Q,\tilde U,\tilde D$), one wino-like triplet fermion ($\tilde W$) and one bino-like singlet fermion ($\tilde B$) with the following quantum numbers,
\begin{eqnarray}
  && 
  \tilde Q = (3,2,1/6),~~~ \tilde U = (3,1,2/3),~~~ \tilde D = (3,1,-1/3),\nonumber\\
  &&
  \tilde W = (1,3,0),~~~ \tilde B = (1,1,0).
\end{eqnarray}
Under this assumption, we can write $\D_2-\D_1$ and $\D_3-\D_1$ as
\begin{eqnarray}
  \D_2-\D_1 &=&
  \frac{1}{15}(7n_{\tilde Q}-4n_{\tilde U}-n_{\tilde D}+20n_{\tilde W}),\\
  \D_3-\D_1 &=&
  \frac{1}{10}(3n_{\tilde Q}-n_{\tilde U}+n_{\tilde D}),
\end{eqnarray}
where the number of the introduced particles are denoted by $n_{\tilde Q},n_{\tilde U},n_{\tilde D}$, and $n_{\tilde W}$ for $\tilde Q,\tilde U,\tilde D,$ and $\tilde W$, respectively~\cite{Giudice:2012zp}. 
Then, we can find a solution for the unification, $n_{\tilde Q}=n_{\tilde U}=n_{\tilde D}=10$ and $n_{\tilde W}=1$, which corresponds to the red point in Fig.~\ref{fig:GCU_map}.
There also exist many possible models that can achieve gauge coupling unification, and this particular choice is one of them.

As we will discuss later, $\tilde W$ is important not only for the gauge coupling unification but also for the colored scalars to decay, although it does not have strong relevance in discussing collider phenomenology of the colored exotics, the scalar $S$ and the $Z'$.
In the case that $\tilde W$ is not lighter than the colored scalars, we need alternative particle so that the colored scalars can decay.
$\tilde B$ can play this role, while it has nothing to do with the gauge coupling unification, which we will also discuss later.

It should be noted that this model requires a large mass splitting in the GUT spectrum, which is similar to the so-called triplet-doublet splitting problem.
Although this splitting suggests a fine-tuning which might need to be addressed eventually, we do not address this as it is likely linked to the triplet-doublet splitting, and just concentrate on the phenomenological aspects.

One important fact about this model is that a rapid proton decay is inevitable due to the tree-level contributions of the $\tilde D$ scalars in a similar way a fast proton decay is mediated by the squarks in the supersymmetric models without the $R$-parity.
One of the convincing symmetries to forbid such a harmful process is the $B-L$ symmetry.
It is natural to consider the $B-L$ as a gauge symmetry in the context of quantum gravity in which any global symmetry might not be respected~\cite{Giddings:1987cg}.
The global $B-L$ symmetry in the SM can be promoted to the local $B-L$ symmetry by introducing three right-handed neutrinos, $N$'s, which are well motivated by the current neutrino data.
We summarize the particle content of the model under consideration in Tab.~\ref{tab:quantum numbers} where $S$ denotes a complex scalar that breaks the $\U(1)_{B-L}$ spontaneously.

Let us comment on the case that the exotic colored particles are fermions.
The purpose of this paper is illustrating a possible connection between the potential diphoton excess at the LHC and a new gauge symmetry by using a specific example.
As we picked the $\U(1)_{B-L}$ for the illustration, we do not need any additional fermionic fields (except for three right-handed neutrinos that are naturally introduced to explain the neutrino mixings) that are usually required for the anomaly cancellations, and we use only the scalar fields $\tilde Q,\tilde U$ and $\tilde D$ for the exotic colored particles.
For any other $\U(1)$ gauge symmetries, additional fermions that are charged under the SM gauge groups are required and some of them can be the colored exotics.
Our approach and analysis, however, basically applies to those more general $\U(1)$ gauge symmetries with a suitable choice of the exotic colored fermions.

\begin{table}
\begin{center}
  \begin{tabular}{c|cccc}
    & $\SU(3)_C$ & $\SU(2)_L$ & $\U(1)_Y$ & $\U(1)_{B-L}$\\
    \hline
    $\tilde Q$'s & 3 & 2 & 1/6 & 1/3\\
    $\tilde U$'s & 3 & 1 & 2/3 & 1/3\\
    $\tilde D$'s & 3 & 1 & $-$1/3 & 1/3\\
    $\tilde W$ & 1 & 3 & 0 & 0\\
    $\tilde B$ & 1 & 1 & 0 & 0\\
    $N$'s & 1 & 1 & 0 & $-$1\\
    $S$ & 1 & 1 & 0 & 2
  \end{tabular}
  \caption{Quantum numbers of the additional particles in the model.}
  \label{tab:quantum numbers}
\end{center}
\end{table}

\begin{figure}[t]
\begin{center}
 \includegraphics[scale=0.7]{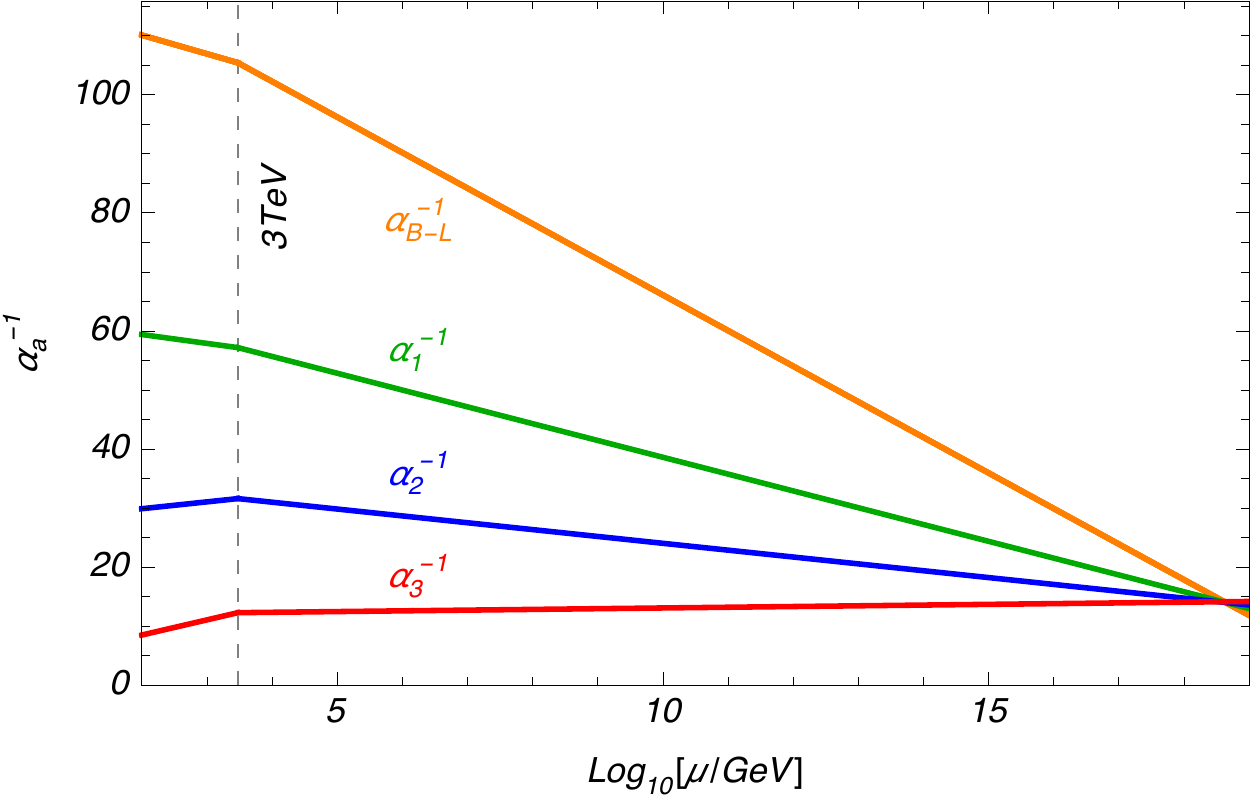}
 \caption{The unification of the SM gauge coupling constants and the $B-L$ gauge coupling constant. The intermediate scale (mass of the exotic particles) is fixed as $M_*=3~\TEV$, and $\m$ denotes the renormalization scale.}
 \label{fig:GCU}
\end{center}
\end{figure}

Once we introduce the $B-L$ gauge symmetry, we have an additional gauge boson, denoted by $Z'$, whose mass is determined by the mass scale of $B-L$ symmetry breaking and the $B-L$ gauge coupling constant at the electroweak scale.
From a viewpoint of the GUT, we expect that the $B-L$ gauge coupling at the electroweak scale is also derived from the universal coupling at the GUT scale.
If we impose the $B-L$ charge on the new particles as shown in Tab.~\ref{tab:quantum numbers}, we obtain $\a_{B-L}^{-1}(M_Z)\simeq 110$, where the coefficient of the beta function of the $B-L$ gauge coupling constant is given by $b_{B-L}=26/3~(\m<M_*)$ and $148/9~(\m>M_*)$ \cite{Basso:2010jm}.
The evolution of the gauge coupling constants is shown in Fig.~\ref{fig:GCU}, where we assume all of the new particles including $N$ and $S$ to be the same mass scale $M_*$, taking $M_*=3~\TEV$ for the illustration.
This particular choice of $M_*$ is not relevant for our result though, and we will take the mass of $S$ as 750 GeV in Sec.~\ref{sec:diphoton}.

It should be noted that $\tilde W$ having mass around 3 TeV can be a good candidate of a dark matter, as it has a similar property of the well-known wino dark matter candidate in the supersymmetric models~\cite{Cirelli:2005uq,Hisano:2006nn}, where the relic abundance is determined by annihilation into the electroweak gauge bosons.
In this case, the neutral component of $\tilde W$ should be slightly lighter than $\tilde Q$'s, $\tilde U$'s and $\tilde D$'s so that it is the lightest stable particle\footnote{Alternatively, the stability of $\tilde W$ can be ensured by imposing $Z_2$ symmetry.}.

Because of the existence of $\tilde B$ in our model, if kinematically allowed, $\tilde U$ and $\tilde D$ can decay into $\tilde B$ and right-handed quarks via dimension four operators in a similar way to the quark-squark-bino interactions in the supersymmetric models.
Thus, $\tilde U$ and $\tilde D$ can decay fast before the big bang nucleosynthesis (BBN) starts.\footnote{
When $\tilde U$ and $\tilde D$ are heavy enough, they can decay before the BBN even without introducing $\tilde B$.
By assuming that $\tilde U$ is slightly larger than $\tilde Q$ and $\tilde W$, $\tilde U$ can decay into $u$, $q$ and $\tilde Q$ via the dimension five operator $M_{\rm GUT}^{-1}(\bar u \tilde U)(\tilde Q^\dagger q)$, where $u$ and $q$ are up-type quark and left-handed quark doublet, respectively.
Since the decay width is given by $\G_{\tilde U}\sim (8\pi)^{-3}(M_{\tilde U}/M_{\rm GUT})^2M_{\tilde U}$ where $M_{\tilde U}$ is the mass of $\tilde U$, the decay temperature becomes
\begin{eqnarray}
	T_D\simeq
	\left(\frac{45}{4\pi^3g_*}M_{\rm PL}^2\G_{\tilde U}^2\right)^{1/4}
	\sim
	2~\MEV\times
	\left(\frac{100}{g_*}\right)^{1/4}
	\left(\frac{M_{\tilde U}}{20~\TEV}\right)^{3/2}
	\left(\frac{10^{16}~\GEV}{M_{\rm GUT}}\right)
\end{eqnarray}
with $g_*$ being the effective number of degrees of freedom.
To obtain this expression, we equate $\G_{\tilde U}$ with $H_D$, the Hubble parameter at $T_D$.
Thus, when the decay temperature satisfies $T_D\gtrsim 1~\MEV$ by taking larger values of $M_{\tilde U}$, the stringent bound from the BBN may be evaded.
The decay of $\tilde D$ is the same as that of $\tilde U$.
The produced $\tilde Q$ immediately decays into $\tilde W$ and $q$ via the interaction of $\bar q \tilde W \tilde Q$ as long as $\tilde W$ is lighter than $\tilde Q$.
}
Depending on the mass, $\tilde B$ or $\tilde W$ can be a dark matter candidate. 
Here, let us briefly comment on the case that $\tilde B$ is the lightest particle.
In the supersymmetric models, it is known that if the bino is the lightest supersymmetric particle, its thermal abundance exceeds the observed dark matter abundance.
On the other hand, coannihilation effect with other superparticles, e.g., wino, can reduce the bino abundance so that the overproduction of bino can be evaded~\cite{Harigaya:2014dwa}.
The situation is similar in our model, and it turns out that $\tilde B$ can be a good candidate of a dark matter if the coannihilation with $\tilde W$ is effective.
In this case, the mass of $\tilde B$ is less than 3 TeV, where the typical mass difference between $\tilde B$ and $\tilde W$ is ${\cal O}(10)~\GEV$.

The potential of scaler fields is given by
\begin{eqnarray}
  &&
  V = V_H(H) + V_S(S) + V_{HS}(H,S) 
  + \sum_{\Phi} [V_\Phi(\Phi)+ V_{H\Phi}(H,\Phi) + V_{S\Phi}(S,\Phi)],\\
  &&
  V_H(H) = -\m_H^2|H|^2 + \lambda_H(|H|^2)^2,~~~
  V_S(S) = -\m_S^2|S|^2 + \lambda_S(|S|^2)^2,\\
  &&
  V_\Phi(\Phi) = \m_\Phi^2|\Phi|^2 + \lambda_\Phi(|\Phi|^2)^2 + \sum_{\Phi'\neq\Phi}\lambda_{\Phi\Phi'}|\Phi|^2|\Phi'|^2,\\
  &&
  V_{HS}(H,S) = \lambda_{HS}|H|^2|S|^2,~~~
  V_{H\Phi}(H,\Phi) = \lambda_{H\Phi}|H|^2|\Phi|^2,\\
  &&
  V_{S\Phi}(S,\Phi) = \lambda_{S\Phi}|S|^2|\Phi|^2,
\end{eqnarray}
where $\Phi$ denotes $\tilde Q$'s, $\tilde U$'s and $\tilde D$'s.
Only the SM Higgs field $H$ and the $B-L$ scalar field $S$ are responsible for the spontaneous breakdown of the electroweak and the $B-L$ symmetries, respectively, and other scalar fields do not have a non-vanishing vacuum expectation value (VEV).
After the symmetry breaking, one can write $S \sim f + (s + i \rho)/\sqrt{2}$, where $s$ and $\rho$ are the real and imaginary part of $S$ in physical basis.
The physical masses of the Higgs boson and $B-L$ scalar are given by
\begin{eqnarray}
  &&
  M_H^2 \simeq \lambda_H v^2 - \frac{(4\lambda_{HS}fv)^2}{\lambda_Sf^2-\lambda_Hv^2},~~~
  M_s^2 \simeq \lambda_S f^2 + \frac{(4\lambda_{HS}fv)^2}{\lambda_Sf^2-\lambda_Hv^2},
  \label{eq:mass}
\end{eqnarray}
where we define the VEVs of $H$ and $S$ as $\langle H\rangle\equiv v$ and $\langle S\rangle\equiv f$, and the mass matrix of $S$ and $H$ is diagonalized by the mixing angle $\tan2\theta=4\l_{HS}fv/(\l_Sf^2-\l_Hv^2)$.
It should be noted that the interactions between the SM Higgs field and the $\Phi$'s affect the Higgs decays such as $H\to\g\g$.
Those effects are, however, constrained by the observed Higgs decays, and thus, $\lambda_{H\Phi}/M_*^2$ is constrained to be small, which does not affect our analysis.
The $\lambda_{HS}$ is also constrained to be small by the bound on the $s \to H H$ mode branching ratio \cite{Franceschini:2015kwy}.

Some of the relevant couplings for the loop correction to the $\lambda_{HS}$ (which are $\lambda_{HS}$ and $\lambda_{H\Phi}$) are set to be sufficiently small at the tree level so that the loop corrected value for this $\lambda_{HS}$ can satisfy the experimental constraints from the aforementioned $s \to H H$ and the $H \to \gamma\gamma$.

There are two possible contributions to the masses of the colored exotics ($\Phi$'s), $M_*^2 = \m_\Phi^2+\lambda_{S\Phi}f^2$, and we take it is dominated by the latter so that the colored exotics have the same scale as the $B-L$ breaking scale.
This common mass scale between the colored exotics and the $B-L$ breaking scale would be rather straightforward when the colored exotics are vector-like fermions, but in our illustration with the colored scalar exotics, it is basically an assumption in the low-energy description of the model.
So we assume the colored exotic masses are given dominantly by the VEV of the $B-L$ scalar $S$,
\begin{eqnarray}
M_*^2\simeq \lambda_{S\Phi}f^2 .
\label{eq:mstar}
\end{eqnarray}

In this model, the new scalars $\Phi$'s couple to the $B-L$ scalar $S$ in the potential term $V_{S\Phi}$.
The interactions among $s$ and $\Phi$'s can be written as
\begin{eqnarray}
  {\cal L} \supset g_{s\Phi} f s\Phi^\dagger \Phi
\end{eqnarray}
by which $s$ can decay into two gauge bosons via loop effects of $\tilde Q$'s, $\tilde U$'s and $\tilde D$'s.
Thus, the relation of Eq.~\eqref{eq:mstar} can be translated into the coupling of the 750 GeV scalar $s$ to the exotic colors as $g_{s\Phi} \equiv \sqrt{2} \lambda_{S \Phi} \simeq \sqrt{2}M_*^2/f^2$.

\section{New diphoton excess}
\label{sec:diphoton}
The decay of the singlet $s$ into $\g\g$ is induced by the loop contributions of $\Phi$'s.
The decay width of this mode is given by
\begin{eqnarray}
  \G_{\g\g} &=&
  \frac{\a^2 M_s^3}{1024\pi^3}
  \left|
  \sum_\Phi \frac{g_{s\Phi}f}{M_*^2}N_\Phi^{\g\g} Q_\Phi^2 A_{0}(\t_\Phi)
  \right|^2,
  \label{eq:G_gamgam}
\end{eqnarray}
where $\a = (3/5) \a_1$, and the sum of $\Phi$ runs over the relevant particles in the loop, and $N_\Phi^{\g\g}$ and $Q_\Phi$ are the number of the particle and its electromagnetic charge, respectively.
In our case, $N_\Phi^{\g\g}$ is given by $N_{\tilde Q}^{\g\g}=6n_{\tilde Q}, N_{\tilde U}^{\g\g}=3n_{\tilde U}, N_{\tilde D}^{\g\g}=3n_{\tilde D}$ with the $n_{\tilde Q}=n_{\tilde U}=n_{\tilde D}=10$.
(When we limit ourselves to the case of the uniform number of copies of the colored exotics, i.e., $n_{\tilde Q}=n_{\tilde U}=n_{\tilde D}$, it turns out our choice of the 10 copies is the only possibility. For detailed discussion, see section~\ref{sec:discussion}.)
$A_{0}(\t_\Phi)$ is the loop function with $\t_\Phi \equiv 4M_s^2/M_*^2$~\cite{Djouadi:2005gi}.\footnote{
The loop function $A_{0}(x)$ is given by 
\begin{eqnarray}
  A_0(x) &=&
  -x^2(2x^{-2}+3x^{-1}+3(2x^{-1}-1)f(x^{-1})),\\
  f(x) &=&
  \left\{
  \begin{array}{lr}
    {\rm arcsin}^2(x^{1/2})&(x\leq1)\\
    -\frac{1}{4}\left[\log\left(\frac{1+\sqrt{1-x}}{1-\sqrt{1-x}}\right)-i\pi\right]^2&(x>1)
  \end{array}
  \right..
\end{eqnarray}
}

\begin{figure}[t]
\begin{center}
 \includegraphics[scale=0.5]{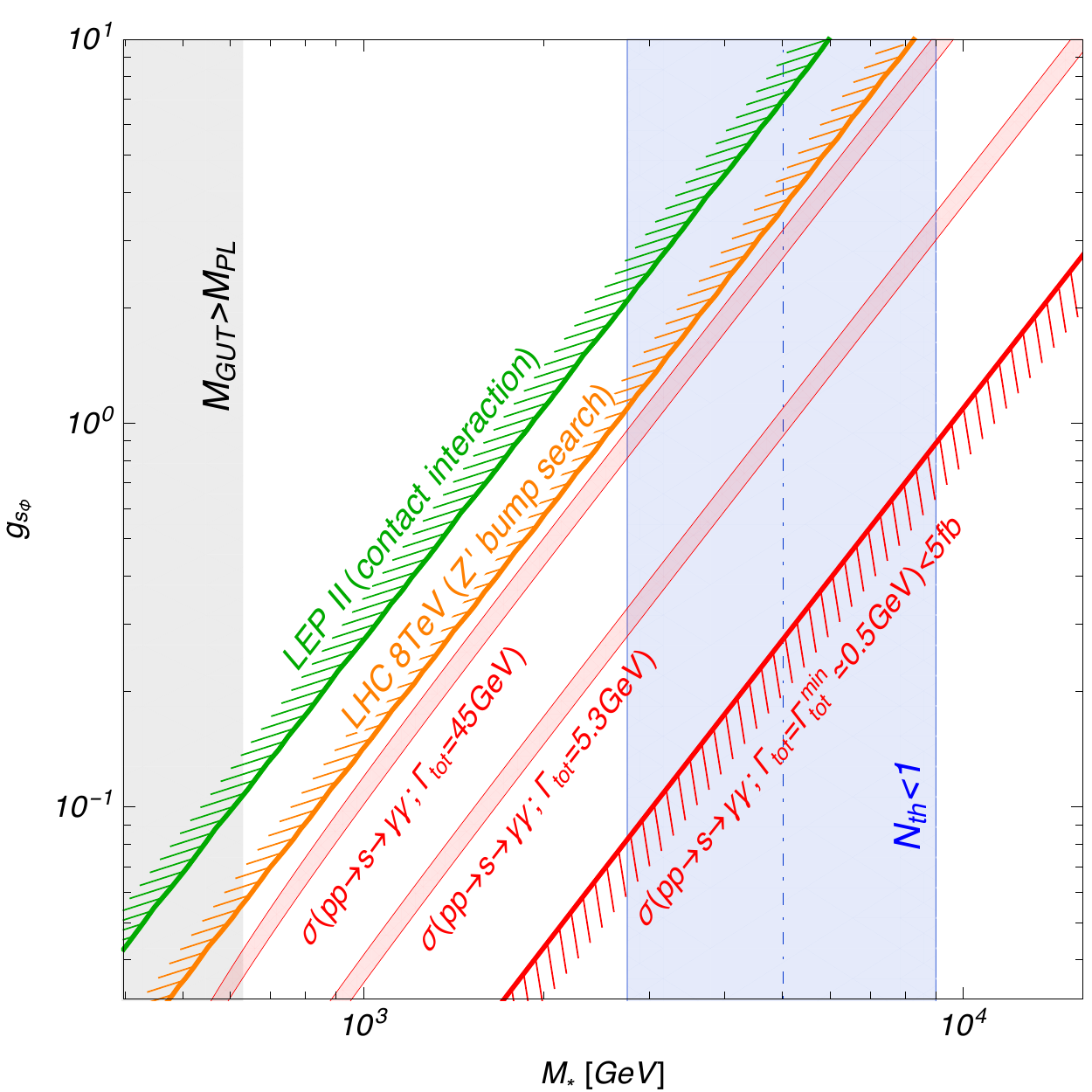}
 \caption{The favored region by the 750 GeV diphoton excess in the $M_*$ and $g_{s\Phi}$ parameter space. The two red bands are the parameter regions that give the cross section $\s(pp\to s\to \g\g)=6\pm 1$ fb for a choice of the total decay width of $\G_{\rm tot}=$ 45 GeV and 5.3 GeV. (See the text for more detail.) The red hatched line corresponds to the $\G_{\rm tot}^{\rm min}$, the minimum decay width from the unavoidable decay channels. The entire parameter region in the panel satisfies $N_{\rm th}<5$, and the very precise gauge coupling unification ($N_{\rm th}<1$) is achieved in the blue region. The light gray region is excluded by requiring $M_{\rm GUT}<M_{\rm PL}$. The constraints on the $Z'$ mass from the LEP II and the LHC are presented by the hatched green line and the hatched orange line, respectively.}
 \label{fig:main1}
\end{center}
\end{figure}

The production of the $s$ is mainly given by gluon-gluon fusion at the LHC.
At the leading order, the cross section of $pp\to s\to \g\g$ is given by
\begin{eqnarray}
  \s(pp\to s\to \g\g) &\simeq&
  \frac{\pi^2}{8}\frac{1}{S}\frac{\G_{gg}}{M_s}{\rm Br}(s\to\g\g)\frac{d{\cal L}_{gg}}{d\t_s},
  \label{eq:sgg}
\end{eqnarray}
where $\G_{gg}$ and ${\rm Br}(s\to\g\g)$ are the decay width of $s\to gg$ and the branching fraction of $s\to\g\g$, and the narrow width approximation is used.
The decay width of $s\to gg$ is given by
\begin{eqnarray}
  \G_{gg} &=&
  \frac{\a_3^2 M_s^3}{512\pi^3}\left|\sum_\Phi N_\Phi^{gg} \frac{g_{s\Phi}f}{M_*^2} A_{0}(\t_\Phi)\right|^2,
  \label{eq:G_gg}
\end{eqnarray}
where $N_\Phi^{gg}$ counts the number of particles of the $\SU(3)_C$ representation in the loop~\cite{Djouadi:2005gi}, where we have $N_{\tilde Q}^{gg}=2n_{\tilde Q}, N_{\tilde U}^{gg}=n_{\tilde U}, N_{\tilde D}^{gg}=n_{\tilde D}$ with the given $n_{\tilde Q}=n_{\tilde U}=n_{\tilde D}=10$.
In Eq.~(\ref{eq:sgg}), $\t_s$ is defined by $\t_s\equiv M_s^2/S$ where $S$ is the squared center-of-mass energy.
The gluon-gluon collision luminosity is given by
\begin{eqnarray}
  \frac{d{\cal L}_{gg}}{d\t_s} &\equiv&
  \int_{-\ln(1/\sqrt{\t_s})}^{\ln(1/\sqrt{\t_s})}dy
  f_{g/p}(\t_s,y)f_{g/p}(\t_s,-y),
\end{eqnarray}
where $f_{g/p}$ is the gluon distribution function in a colliding proton, and $y$ is the rapidity of the colliding gluon-gluon system.
In a numerical analysis, we utilize the distribution function of MSTW 2008lo~\cite{Martin:2009iq}.

The combined analysis of the ATLAS and CMS gave the cross section $\s(pp\to s\to\g\g)\simeq 6\pm 1$ fb (95\% CL) in 2015 \cite{Ellis:2015oso}.
We show the favored region by the $\s(pp\to s\to \g\g)$ measurement in $M_*$ and $g_{s\Phi}$ parameter space in Fig.~\ref{fig:main1}, which is depicted by the red bands.
The two red bands correspond to the $\G_{\rm tot} = 45~\GEV$ and $5.3~\GEV$, which corresponds to the fitted total decay width in ATLAS and the diphoton invariant mass resolution in ATLAS, respectively, reported in 2015~\cite{ATLAS}.
Our model has the minimum width ($\G_{\rm tot}^{\rm min}$) for the 750 GeV scalar $s$ from the unavoidable decay channels of $s\to gg/\g\g/WW/ZZ$ and $Z\g$, which is approximately $\G_{\rm tot}^{\rm min}\simeq 0.5~\GEV$ when the cross section of the 750 GeV diphoton excess is explained, and it corresponds to the hatched red line in Fig.~\ref{fig:main1}.
The total decay width depends on the unknown decay modes as well
\footnote{There exist other possible decay channels such as $s\to NN$ which will be discussed in Sec.~\ref{sec:other experimental constraints}.
  The minimum decay width, $\G_{\rm tot}^{\rm min}$ means the total width obtained by turning off these channels.
  This can be realized by, for instance, taking the mass of $N$ larger than $M_s / 2$.
}, which are not fixed yet and can be considered arbitrary at this stage.
We will discuss more about the 750 GeV scalar decay width in Sec.~\ref{sec:other experimental constraints}.

The GUT scale is fixed by only $M_*$, and we put a lower bound, $M_*\gtrsim 630~\GEV$, by requiring $M_{\rm GUT}<M_{\rm PL}$, which is shown in the light gray region.
The bounds from squark searches are also equivalent to the bounds on $M_*$.
The current sbottom searches at the LHC experiments give $M_*\gtrsim700~\GEV$ provided that the lightest supersymmetric particle is $O(100)~\GEV$~\cite{Khachatryan:2015wza}, and the diphoton excess can be explained in most of the parameter region of our model.
In all parameter region shown in Fig.~\ref{fig:main1}, the accuracy of the coupling unification is within the range of $N_{\rm th}<5$, and we show the very precise gauge coupling unification region ($N_{\rm th}<1$), which is the blue region.

\section{Implications for a new heavy gauge boson}
\label{sec:implications for new gauge boson}
\begin{figure}[tbh]
\begin{center}
 \includegraphics[scale=0.5]{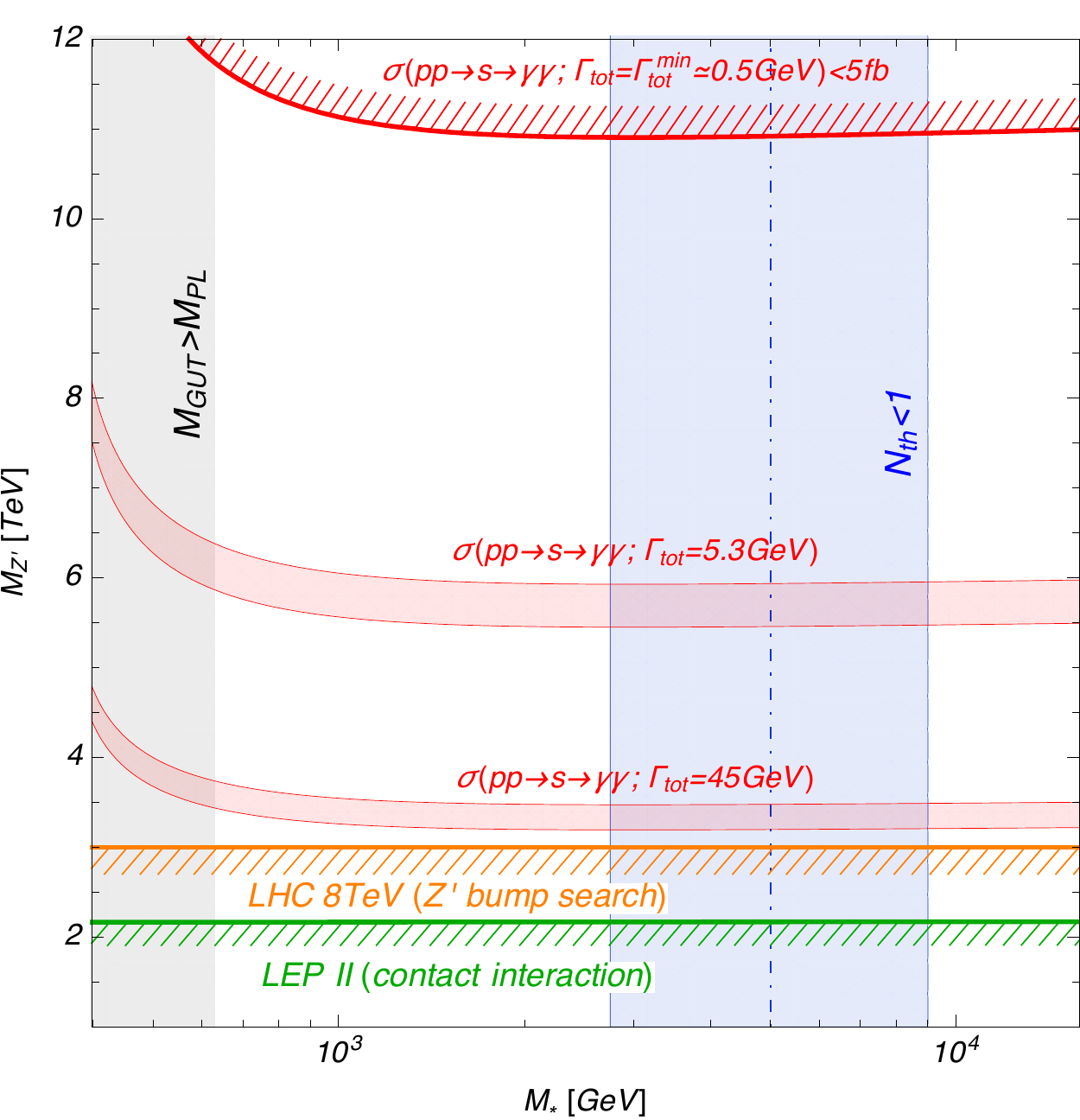}
 \hspace*{12mm}
 \includegraphics[scale=0.5]{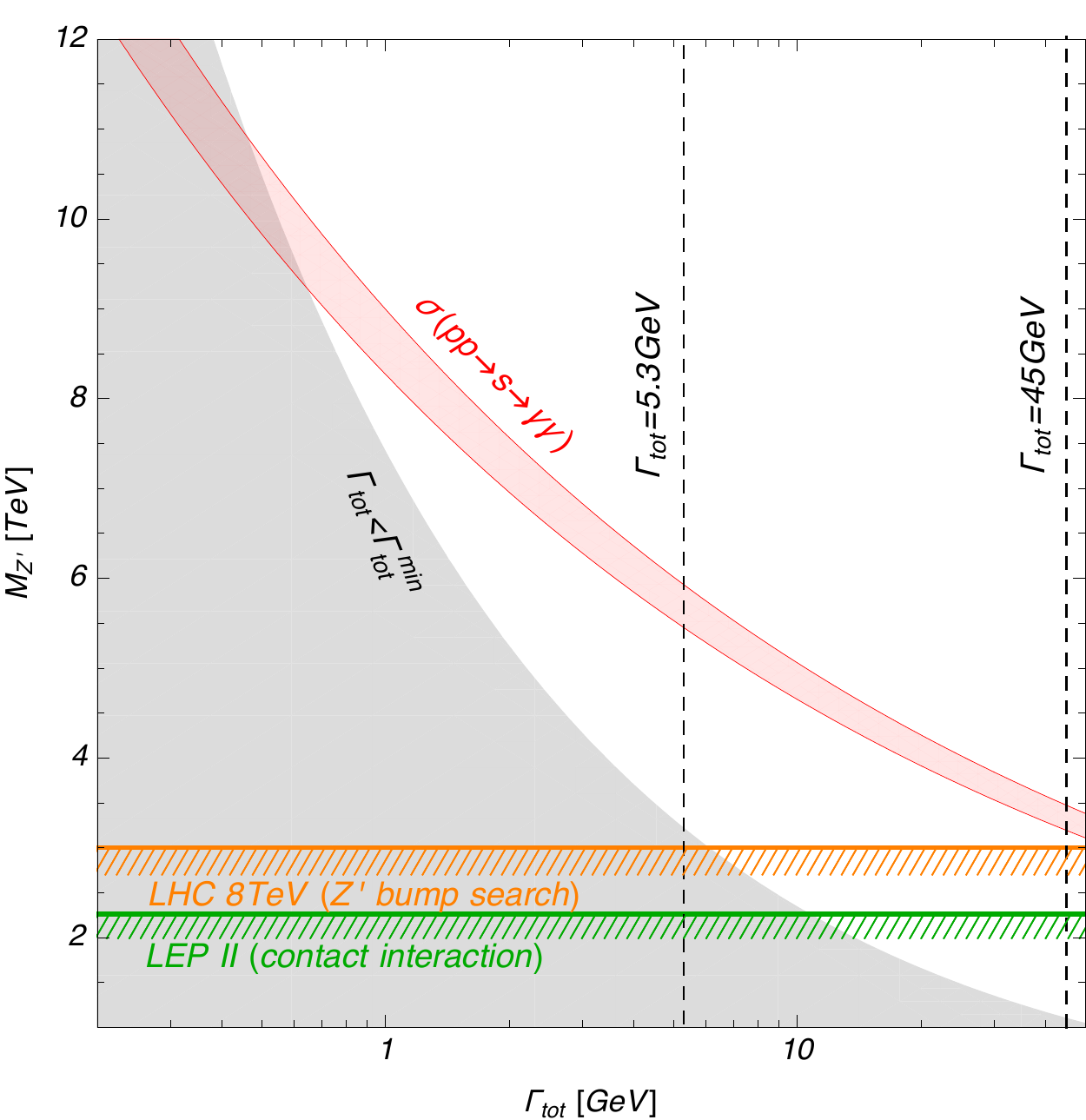} \\
 \caption{The left panel is a translation of Fig.~\ref{fig:main1} to the $M_*$ and $M_{Z'}$ parameter space.
The favored region for the 750 GeV cross section $\s(pp\to s\to \g\g)=6\pm 1$ fb with the total width being $\G_{\rm tot}=$ 45 GeV, 5.3 GeV is shown in the red band. The predicted $M_{Z'}$ is not sensitive to the $M_*$ as long as $M_* \ge 1~\TEV$ (flat region). In the blue region, very precise gauge coupling unification ($N_{\rm th} < 1$) is achieved. The right panel shows how the $M_{Z'}$ prediction varies with $\G_{\rm tot}$ for the flat band region ($M_* \ge 1~\TEV$) with a specific choice of $M_* = 3~\TEV$. The dark gray region in the right panel is eliminated by $\G_{\rm tot}<\G_{\rm tot}^{\rm min}$.}
 \label{fig:main}
\end{center}
\end{figure}

The gauged $\U(1)_{B-L}$ symmetry in our model leads to the existence of a new gauge boson, $Z'$, whose mass is predicted by the condition of the gauge coupling unification and the diphoton signal at the LHC.
The mass of $Z'$ is generated via the spontaneous $B-L$ breaking, and hence
\begin{eqnarray}
  M_{Z'}^2 &=& 8g_{B-L}^2f^2, \label{eq:MZprime}
\end{eqnarray}
where $g_{B-L}^2=4\pi\a_{B-L}$ is the $\U(1)_{B-L}$ gauge coupling constant.
As discussed in Sec.~\ref{sec:the model}, if $\a_{B-L}$ is also unified at the GUT scale, the low energy value of $\a_{B-L}$ can be derived from the universal coupling constant $\a_U(M_{\rm GUT})$ which is a function of $M_*$ in our setup.
On the other hand, the $B-L$ breaking scale $f$ is related to the coupling $g_{s\Phi}$.
Therefore, if the diphoton signal is confirmed, the scale of $Z'$ can also be predicted in our model.

In the left panel of Fig.~\ref{fig:main}, we show $\s(pp\to s\to\g\g)$ as a function of $M_*$ and $M_{Z'}$.
As the figure shows, the $M_*$-dependence of the predicted $M_{Z'}$ is weak because the TeV-scale $M_*$ does not change $\a_{B-L}$ at the electroweak scale sensitively.
This weak dependence even practically vanishes for $M_* \ge 1~\TEV$.
The very precise gauge coupling unification (blue region) also corresponds to this case.

The LEP-II experiment gave the bound on the ratio of $Z'$ mass and the coupling constant, $M_{Z'}/g_{B-L}\gtrsim6.7~\TEV$ using a contact interaction~\cite{Carena:2004xs}, which is shown by the green hatched line.
Currently, the LHC 8 TeV has set more stringent constraint on the $Z'$ mass as $M_{Z'}\gtrsim3~\TEV$~\cite{Chatrchyan:2012oaa} (see also Ref.~\cite{Guo:2015lxa}) using a dilepton bump search, which is presented by the orange hatched line.
It should be noted that the LHC gives the bound on the $(M_{Z'},\sigma(pp\to Z')\cdot {\rm Br}(Z'\to l^+l^-))$ plane, where $\sigma(pp\to Z')$ and ${\rm Br}(Z'\to l^+l^-)$ are the production cross section of $Z'$ and its branching fraction into dilepton, respectively.
The LHC bound in the figure does not include the $Z'$ decay into the colored scalars.
In the case of $2M_*\lesssim M_{Z'}$, however, the decay into pairs of the colored scalars is kinematically allowed, by which the bound would get relaxed \footnote{The LEP-II bound may also be slightly changed as the $Z'$ total decay width increases.}, while it does not affect our main results.
In the future LHC experiments, while collider signatures of such new decay modes strongly depend on the decay of the colored scalars, it is interesting to investigate collider phenomenology of this channel.

In order to explain the 750 GeV diphoton cross section, a smaller total decay width requires a larger $B-L$ breaking scale.
This is because the proportionality of the diphoton cross section is approximately given by
\begin{eqnarray}
\s(pp\to s\to\g\g) \propto \frac{\G_{gg}\G_{\g\g}}{M_s\G_{\rm tot}} \propto \frac{M_s^5}{f^4\G_{\rm tot}} \, ,
\end{eqnarray}
and thus $f$ and $\G_{\rm tot}$ need to balance with each other so that the 750 GeV cross section can be achieved.
This leads to the $\G_{\rm tot}$-dependence of the $M_{Z'}$, which is shown in the right panel of Fig.~\ref{fig:main}. 
By using $\a_{B-L}^{-1}(M_Z)\sim 110$ of Fig.~\ref{fig:GCU}, the VEV $f$ can be obtained from the $M_{Z'}$ as $f\sim 0.96 ~ M_{Z'}$ from Eq.~\eqref{eq:MZprime}.
While we take $M_*=3~\TEV$ for definiteness in the right panel of Fig.~\ref{fig:main}, the result remains almost the same for $M_* \ge 1~\TEV$ or when the gauge coupling unification is very precise ($N_{\rm th}<1$) as is clear from the left panel of Fig.~\ref{fig:main}.
The dark gray region in the right panel of Fig.~\ref{fig:main} is excluded because it results in $\G_{\rm tot}<\G_{\rm tot}^{\rm min}$.
Therefore, $M_{Z'}$ is predicted to be in the range of $3.5~\TEV\lesssim M_{Z'}\lesssim 11~\TEV$ in our model in order to satisfy the cross section of the 750 GeV scalar.
The lower bound on the $M_{Z'}$ ($\G_{\rm tot} = 45~\GEV$ case) is very close to the bound from the dilepton resonance search with the LHC 8 TeV data (orange hatched line).
Considering the assumed width is actually rather large, when we require the width to be at least the $5.3~\GEV$, we get $3.5~\TEV\lesssim M_{Z'}\lesssim 5.8~\TEV$, which is expected to be covered by the LHC Run 2 experiments~\cite{Godfrey:2013eta}.

A possible connection between the 750 GeV diphoton excess and the $Z'$ that we established in this work also implies interesting decay modes of the $Z'$.
For instance, the decay channel induced by loop contributions of the exotic colored particles $\Phi$,
$Z'\to s\gamma / sZ$ followed by $s\to gg$ will be a direct indication that the 750 GeV diphoton excess is linked to the TeV-scale $Z'$.
There are other important decay modes such as $Z' \to s\gamma/sZ$ followed by $s \to NN$, which is relevant for the relatively large $\G_{\rm tot}$ case (see the following section).

Before closing this section, let us discuss an influence of a non-zero $\m_\Phi$ (the bare mass term of the squark-like particles) on our results.
It appears only through the relation of Eq.~(\ref{eq:mstar}) where we have assumed that the masses of the colored exotics dominantly come from the VEV of $S$.
Now, we discuss how much the prediction of the $M_{Z'}$  can be affected by a non-zero value of the $\m_\Phi$.
By using $\d\equiv \m_\Phi^2/(\l_{S\Phi}f^2)$, the colored exotic masses are given by 
\begin{eqnarray}
  M_*^2=\l_{S\Phi}f^2(1+\d)
  \label{eq:Mstar2}
\end{eqnarray} 
where $\d=0$ leads to our predictions for the $M_{Z'}$ in Fig.~\ref{fig:main}.
As discussed in this section, the scale $f$ can be obtained from the cross section $\s(pp\to s\to \g\g)$ because of its proportionality $\s(pp\to s\to \g\g)\propto \G_{gg}\G_{\g\g}\sim (1/f^2)^2$.
By substituting Eq.~(\ref{eq:Mstar2}) into Eqs.~(\ref{eq:G_gamgam}) and (\ref{eq:G_gg}), we now obtain
\begin{eqnarray}
  \G_{\g\g}, \G_{gg}
  &\propto&
  \left|\frac{g_{s\Phi}f}{M_*^2}\right|^2
  =\frac{2}{f^2(1+\d)^2}
  \equiv \frac{2}{f'^{2}},
\end{eqnarray}
and thus, we have $\s(pp\to s\to \g\g)\propto \G_{gg}\G_{\g\g}\sim (1/f'^2)^2$.
The $f'$ can be obtained from the 750 GeV diphoton cross section, and it serves as a upper bound on the $f$.
In other words, our results for $\d=0$ ($M_{Z'}|_{\d=0}$) in Fig.~\ref{fig:main} can be translated into those for $\d\neq0$ ($M_{Z'}|_{\d\neq0}$) by using $M_{Z'}|_{\d\neq0}=M_{Z'}|_{\d=0}/ (1+\d)$ from Eq.~\eqref{eq:MZprime}.
Now, the result of Fig.~\ref{fig:main} can be understood as the upper bound on the $M_{Z'}$, which is the $M_{Z'}$ itself when the bare mass term is zero.

\section{Other experimental constraints}
\label{sec:other experimental constraints}
Since our model involves the $\SU(2)_L$ and/or $\U(1)_Y$ charged particles, $s$ can also decay into electroweak gauge bosons such as $WW, ZZ$ and $Z\g$.
Being consistent with our approach of taking the late 2015 situation as our example, we use the extrapolated values from the limit of 8 TeV to that of 13 TeV~\cite{Franceschini:2015kwy}.

In our case, the ratios of the decay widths (see Ref.~\cite{Craig:2015lra}) are
\begin{eqnarray}
  &&
  \frac{\G(s\to WW)}{\G(s\to \g\g)} \sim 7 < 20,~~~
  \frac{\G(s\to ZZ)}{\G(s\to \g\g)} \sim 3 < 6,~~~
  \frac{\G(s\to Z\g)}{\G(s\to \g\g)} \sim 0.5 < 2,
\end{eqnarray}
where the rightmost side in each equation shows the extrapolated values by assuming the production cross section at 13 TeV is about five times larger than that at 8 TeV, which is reasonable if the production is mainly through the gluon fusion.

As reported in Refs.~\cite{ATLAS,CMS}, the best fit value of the total width of the 750 GeV excess was 45 GeV.
Let us briefly discuss this point.
In our setup, the 750 GeV scalar $s$ can decay into three pairs of right-handed neutrinos via the interaction of $(\kappa/2)S\overline{N^C}_iN_i$ ($i=1,2,3$) with the partial width
\begin{eqnarray}
	\sum_{i=1}^{3}\G(s\to N_iN_i) 
	= \frac{3\lambda_S}{8\pi^2}\left(\frac{M_N}{M_s}\right)^2\left(1-\frac{4M_N^2}{M_s^2}\right)^{3/2},
\end{eqnarray}
where we have used $M_s^2=2\lambda_Sf^2$ and a common Majorana mass $M_N=\kappa f$.
When the right-handed neutrino mass is $M_N^2=M_s^2/10\sim(237~\GEV)^2$, the decay width can be maximally large, $\sum_{i=1}^{3}\G(s\to N_iN_i)\sim 4.2\times\lambda_S$.
Therefore, a sizable coupling of the $\lambda_S$ may realize the large decay width.
All three pairs of the produced $N$ can decay into $\ell W^{\pm}$ with $W^{\pm}\to jj/\ell \nu$, which can be investigated in future collider experiments.
This large width of the 750 GeV scalar predicts the relatively light $Z'$, $M_{Z'} \simeq 3.5 ~\TEV$, which is only slightly above the current bound on the $Z'$ mass.
In this scenario, the discovery of the $Z'$ at the LHC Run 2 may occur very soon.

\section{Discussion on the possible models}
\label{sec:discussion}
There potentially exist many possibilities in choosing the exotic particle contents, besides the one we chose in section~\ref{sec:the model}, that can account for the potential diphoton excess.
Here let us briefly look at the other choices of the exotic particle contents.

Once a certain value of $M_*$ is fixed, we obtain $N_\Phi^{\g\g}Q_\Phi^2$, $N_\Phi^{gg}$ and $\a_{B-L}(M_Z)$ by requiring the gauge coupling unification.
Then, the cross section $\s(pp\to s\to \g\g)$ can be written as a function of $f$, which is determined by taking an appropriate value of $\s(pp\to s\to \g\g)$ (and also the total width).
As a result, $M_{Z'}$ is predicted for a given model.

As a matter of fact, it turns out, with the $M_* = 3~\TEV$ motivated from the dark matter relic density constraint, when we assume the same number of copies for the colored exotics, i.e., $n_{\tilde q} \equiv n_{\tilde Q}=n_{\tilde U}=n_{\tilde D}$, and take the gauge coupling unification threshold $N_{\rm th}<5$, our choice of $n_{\tilde q} = 10$ in section~\ref{sec:the model} is the only viable model that can explain the diphoton excess of the 750 GeV example while satisfying all relevant experimental constraints.
(The $n_{\tilde q} = 9$ case predicts $M_{Z'}\approx 3.0~\TEV$, which is almost the exact bound given by the CMS search only for the $Z'$ \cite{Chatrchyan:2012oaa,Guo:2015lxa}. Since the experimental bound should be stronger when it is combined with the ATLAS result, we take this case is excluded.)

If we relax the condition for the $N_{\rm th}$, e.g., $N_{\rm th}<10$, the diphoton excess can be also explained by models with the smaller $n_{\tilde q}$ ($n_{\tilde q} =6 - 8$).
The scale of $f$ in these models, however, needs to be smaller than that of the $n_{\tilde q}=9$ case in order to explain the diphoton excess, which predicts $M_{Z'}<3~\TEV$, and get excluded by the data.
Therefore, the relaxation of the $N_{\rm th}$ does not change our conclusion of this paper at a meaningful level.

It should be noted that if we take $n_{\tilde Q}$, $n_{\tilde U}$ and $n_{\tilde D}$ independently, we have more variety of models which can explain the diphoton excess.
To get a viable $M_{Z'}$ ($>3~\TEV$), we have checked that $n_{\tilde Q} \ge 9$ and $n_{\tilde U} \ge 9$ is needed in this case, e.g., $(n_{\tilde Q}, n_{\tilde U}, n_{\tilde D}) = (9,9,10)$, $(9,10,8)$, $(9,11,6)$.
While there may exist many possible choices when we consider this non-universal number of copies for the colored exotics, we do not pursue the analysis of this diversity since our aim of this paper is to illustrate a possible connection between the potential diphoton excess and the $Z'$ with a working example.

Consequently, we can state that our analysis and conclusions with the choice of $n_{\tilde q} = 10$ is rather generic (as long as the uniform number of copies of the colored exotics is considered), and it is fairly insensitive to the GUT threshold $N_{\rm th} < 5$.

\section{Summary}
\label{sec:summary}
One of the most natural explanations of the potential diphoton excess signals at the LHC experiments, like the recently highlighted 750 GeV case, could be a new scalar boson that couples to the exotic $\SU(3)_C\times\U(1)_{em}$ charged particles.
Although those particles indicate new physics beyond the SM, the role of them in nature is unclear.
One possibility is that the new scalar boson is responsible for the breakdown of a new gauge symmetry, and gives a mass to a new gauge boson, $Z'$.

In this paper, we studied implications for the $Z'$ of the potential diphoton excess with a specific example of the 750 GeV data of late 2015 \cite{ATLAS,CMS}.
We discussed the scenario using a popular $B-L$ gauge symmetry together with the idea of gauge coupling unification.
In particular, we investigated the model including squark-like particles, which not only realize the gauge coupling unification at the GUT scale, but also enable the new 750 GeV scalar $s$ to be produced by the gluon fusion and to decay into diphoton.
A large number of colored exotics are necessary in this scenario in order to fit the 750 GeV diphoton cross section and to satisfy the current $Z'$ mass bound.
We find the model with ten copies of $\tilde Q, \tilde U$ and $\tilde D$ viable.
We assume the scalar is also responsible to the masses of these colored scalar particles.
The $B-L$ is well motivated in this scenario as it prevents too a fast proton decay mediated by the squark-like particles just as the $R$-parity or its gauge origin $B-L$ does the same task in the supersymmetric models.
It is natural to consider that the $B-L$ gauge coupling constant is also unified at the GUT scale, and this expectation allows us to predict the value of the $B-L$ gauge coupling constant at low energies.
On the other hand, the cross section of the $pp\to s\to\g\g$ is sensitive to the scale of the $\U(1)$ symmetry breaking.
By putting both the gauge coupling unification and the diphoton signal together, we found that the mass of the $Z'$ in our scenario is expected to be slightly beyond the current limit.
In a more general case, in which the colored exotic scalars get the masses not only from the 750 GeV scalar, but also from the non-zero bare mass terms, our results for the $Z'$ mass serves as an upper bound.
Thus, the $Z'$ in this scenario is likely to be discovered at the LHC Run 2 experiments.

As our study illustrates, it is important to keep it in mind that a discovery of a new scalar boson would suggests a possible new gauge boson of the same scale which can be searched for at the LHC experiments. 
Thus the search for a new gauge boson at the LHC experiments would be highly motivated afterwards. \\

\section*{Acknowledgements}
This work was supported by IBS under the project code, IBS-R018-D1.


\begin{thebibliography}{99}  
\bibitem{ATLAS}
ATLAS note, ATLAS-CONF-2015-081.
\bibitem{CMS}
CMS note, CMS PAS EXO-15-004.

\bibitem{Moriond2016}
M.~Delmastro [ATLAS Collaboration],
Diphoton searches in ATLAS,
Les Rencontres de Moriond EW 2016 (March 17, 2016);
P.~Musella [CMS Collaboration],
Search for high mass diphoton resonances at CMS,
Les Rencontres de Moriond EW 2016 (March 17, 2016);
ATLAS note CONF-2016-018;
CMS note PAS EXO-16-018.

\bibitem{ICHEP}
  The ATLAS collaboration [ATLAS Collaboration],
  ATLAS-CONF-2016-059;
  CMS Collaboration [CMS Collaboration],
  CMS-PAS-EXO-16-027.

\bibitem{diphoton:composite/NGB}
  K.~Harigaya and Y.~Nomura,
  arXiv:1512.04850 [hep-ph];
  Y.~Nakai, R.~Sato and K.~Tobioka,
  arXiv:1512.04924 [hep-ph];
  A.~Pilaftsis,
  arXiv:1512.04931 [hep-ph];
  T.~Higaki, K.~S.~Jeong, N.~Kitajima and F.~Takahashi,
  arXiv:1512.05295 [hep-ph];
  M.~Low, A.~Tesi and L.~T.~Wang,
  arXiv:1512.05328 [hep-ph];
  B.~Bellazzini, R.~Franceschini, F.~Sala and J.~Serra,
  arXiv:1512.05330 [hep-ph];
  C.~Petersson and R.~Torre,
  arXiv:1512.05333 [hep-ph];
  E.~Molinaro, F.~Sannino and N.~Vignaroli,
  arXiv:1512.05334 [hep-ph];
  S.~Matsuzaki and K.~Yamawaki,
  arXiv:1512.05564 [hep-ph];
  J.~M.~No, V.~Sanz and J.~Setford,
  arXiv:1512.05700 [hep-ph];
  S.~V.~Demidov and D.~S.~Gorbunov,
  arXiv:1512.05723 [hep-ph];
  L.~Bian, N.~Chen, D.~Liu and J.~Shu,
  arXiv:1512.05759 [hep-ph];
  J.~S.~Kim, J.~Reuter, K.~Rolbiecki and R.~R.~de Austri,
  arXiv:1512.06083 [hep-ph];
  M.~T.~Arun and P.~Saha,
  arXiv:1512.06335 [hep-ph];
  W.~Liao and H.~q.~Zheng,
  arXiv:1512.06741 [hep-ph];
  L.~Berthier, J.~M.~Cline, W.~Shepherd and M.~Trott,
  arXiv:1512.06799 [hep-ph];
  J.~M.~Cline and Z.~Liu,
  arXiv:1512.06827 [hep-ph];
  J.~A.~Casas, J.~R.~Espinosa and J.~M.~Moreno,
  arXiv:1512.07895 [hep-ph].

\bibitem{diphoton:extra dimension}
  P.~Cox, A.~D.~Medina, T.~S.~Ray and A.~Spray,
  arXiv:1512.05618 [hep-ph];
  A.~Ahmed, B.~M.~Dillon, B.~Grzadkowski, J.~F.~Gunion and Y.~Jiang,
  arXiv:1512.05771 [hep-ph];
  E.~Megias, O.~Pujolas and M.~Quiros,
  arXiv:1512.06106 [hep-ph];
  M.~T.~Arun and P.~Saha,
  arXiv:1512.06335 [hep-ph];
  D.~Bardhan, D.~Bhatia, A.~Chakraborty, U.~Maitra, S.~Raychaudhuri and T.~Samui,
  arXiv:1512.06674 [hep-ph];
  J.~J.~Heckman,
  arXiv:1512.06773 [hep-ph];
  H.~Davoudiasl and C.~Zhang,
  arXiv:1512.07672 [hep-ph];
  C.~Cai, Z.~H.~Yu and H.~H.~Zhang,
  arXiv:1512.08440 [hep-ph];
  Q.~H.~Cao, Y.~Liu, K.~P.~Xie, B.~Yan and D.~M.~Zhang,
  arXiv:1512.08441 [hep-ph];
  L.~A.~Anchordoqui, I.~Antoniadis, H.~Goldberg, X.~Huang, D.~Lust and T.~R.~Taylor,
  arXiv:1512.08502 [hep-ph].

\bibitem{diphoton:additional charged particles}
  A.~Angelescu, A.~Djouadi and G.~Moreau,
  arXiv:1512.04921 [hep-ph];
  S.~Knapen, T.~Melia, M.~Papucci and K.~Zurek,
  arXiv:1512.04928 [hep-ph];
  D.~Buttazzo, A.~Greljo and D.~Marzocca,
  arXiv:1512.04929 [hep-ph];
  S.~Di Chiara, L.~Marzola and M.~Raidal,
  arXiv:1512.04939 [hep-ph];
  S.~D.~McDermott, P.~Meade and H.~Ramani,
  arXiv:1512.05326 [hep-ph];
  R.~S.~Gupta, S.~Jäger, Y.~Kats, G.~Perez and E.~Stamou,
  arXiv:1512.05332 [hep-ph];
  A.~Kobakhidze, F.~Wang, L.~Wu, J.~M.~Yang and M.~Zhang,
  arXiv:1512.05585 [hep-ph];
  W.~Chao, R.~Huo and J.~H.~Yu,
  arXiv:1512.05738 [hep-ph];
  S.~Fichet, G.~von Gersdorff and C.~Royon,
  arXiv:1512.05751 [hep-ph];
  D.~Curtin and C.~B.~Verhaaren,
  arXiv:1512.05753 [hep-ph];
  P.~Agrawal, J.~Fan, B.~Heidenreich, M.~Reece and M.~Strassler,
  arXiv:1512.05775 [hep-ph];
  C.~Csaki, J.~Hubisz and J.~Terning,
  arXiv:1512.05776 [hep-ph];
  A.~Falkowski, O.~Slone and T.~Volansky,
  arXiv:1512.05777 [hep-ph];
  D.~Aloni, K.~Blum, A.~Dery, A.~Efrati and Y.~Nir,
  arXiv:1512.05778 [hep-ph];
  E.~Gabrielli, K.~Kannike, B.~Mele, M.~Raidal, C.~Spethmann and H.~Veermäe,
  arXiv:1512.05961 [hep-ph];
  R.~Benbrik, C.~H.~Chen and T.~Nomura,
  arXiv:1512.06028 [hep-ph];
  A.~Alves, A.~G.~Dias and K.~Sinha,
  arXiv:1512.06091 [hep-ph];
  L.~M.~Carpenter, R.~Colburn and J.~Goodman,
  arXiv:1512.06107 [hep-ph];
  I.~Chakraborty and A.~Kundu,
  arXiv:1512.06508 [hep-ph];
  R.~Ding, L.~Huang, T.~Li and B.~Zhu,
  arXiv:1512.06560 [hep-ph];
  H.~Han, S.~Wang and S.~Zheng,
  arXiv:1512.06562 [hep-ph];
  M.~x.~Luo, K.~Wang, T.~Xu, L.~Zhang and G.~Zhu,
  arXiv:1512.06670 [hep-ph];
  O.~Antipin, M.~Mojaza and F.~Sannino,
  arXiv:1512.06708 [hep-ph];
  F.~Wang, L.~Wu, J.~M.~Yang and M.~Zhang,
  arXiv:1512.06715 [hep-ph];
  J.~Cao, C.~Han, L.~Shang, W.~Su, J.~M.~Yang and Y.~Zhang,
  arXiv:1512.06728 [hep-ph];
  M.~Dhuria and G.~Goswami,
  arXiv:1512.06782 [hep-ph];
  S.~M.~Boucenna, S.~Morisi and A.~Vicente,
  arXiv:1512.06878 [hep-ph];
  C.~W.~Murphy,
  arXiv:1512.06976 [hep-ph];
  G.~M.~Pelaggi, A.~Strumia and E.~Vigiani,
  arXiv:1512.07225 [hep-ph];
  K.~M.~Patel and P.~Sharma,
  arXiv:1512.07468 [hep-ph];
  S.~Chakraborty, A.~Chakraborty and S.~Raychaudhuri,
  arXiv:1512.07527 [hep-ph];
  M.~Cvetic, J.~Halverson and P.~Langacker,
  arXiv:1512.07622 [hep-ph];
  B.~C.~Allanach, P.~S.~B.~Dev, S.~A.~Renner and K.~Sakurai,
  arXiv:1512.07645 [hep-ph];
  K.~Cheung, P.~Ko, J.~S.~Lee, J.~Park and P.~Y.~Tseng,
  arXiv:1512.07853 [hep-ph];
  J.~Zhang and S.~Zhou,
  arXiv:1512.07889 [hep-ph];
  L.~J.~Hall, K.~Harigaya and Y.~Nomura,
  arXiv:1512.07904 [hep-ph];
  A.~Salvio and A.~Mazumdar,
  arXiv:1512.08184 [hep-ph];
  D.~Chway, R.~Dermíšek, T.~H.~Jung and H.~D.~Kim,
  arXiv:1512.08221 [hep-ph];
  G.~Li, Y.~n.~Mao, Y.~L.~Tang, C.~Zhang, Y.~Zhou and S.~h.~Zhu,
  arXiv:1512.08255 [hep-ph];
  M.~Son and A.~Urbano,
  arXiv:1512.08307 [hep-ph];
  H.~An, C.~Cheung and Y.~Zhang,
  arXiv:1512.08378 [hep-ph];
  J.~Cao, F.~Wang and Y.~Zhang,
  arXiv:1512.08392 [hep-ph];
  F.~Wang, W.~Wang, L.~Wu, J.~M.~Yang and M.~Zhang,
  arXiv:1512.08434 [hep-ph];
  W.~Chao,
  arXiv:1512.08484 [hep-ph];
  P.~S.~B.~Dev, R.~N.~Mohapatra and Y.~Zhang,
  arXiv:1512.08507 [hep-ph].

\bibitem{diphoton:extended gauge sector}
  R.~Martinez, F.~Ochoa and C.~F.~Sierra,
  arXiv:1512.05617 [hep-ph];
  W.~Chao,
  arXiv:1512.06297 [hep-ph];
  S.~Chang,
  arXiv:1512.06426 [hep-ph];
  T.~F.~Feng, X.~Q.~Li, H.~B.~Zhang and S.~M.~Zhao,
  arXiv:1512.06696 [hep-ph];
  S.~M.~Boucenna, S.~Morisi and A.~Vicente,
  arXiv:1512.06878 [hep-ph];
  A.~E.~C.~Hernández and I.~Nisandzic,
  arXiv:1512.07165 [hep-ph];
  J.~de Blas, J.~Santiago and R.~Vega-Morales,
  arXiv:1512.07229 [hep-ph];
  Q.~H.~Cao, S.~L.~Chen and P.~H.~Gu,
  arXiv:1512.07541 [hep-ph];
  K.~Das and S.~K.~Rai,
  arXiv:1512.07789 [hep-ph];
  J.~Liu, X.~P.~Wang and W.~Xue,
  arXiv:1512.07885 [hep-ph].

\bibitem{diphoton:extended Higgs sector}
  X.~F.~Han and L.~Wang,
  arXiv:1512.06587 [hep-ph];
  J.~Chang, K.~Cheung and C.~T.~Lu,
  arXiv:1512.06671 [hep-ph];
  W.~C.~Huang, Y.~L.~S.~Tsai and T.~C.~Yuan,
  arXiv:1512.07268 [hep-ph];
  S.~Moretti and K.~Yagyu,
  arXiv:1512.07462 [hep-ph];
  M.~Badziak,
  arXiv:1512.07497 [hep-ph];
  X.~J.~Bi {\it et al.},
  arXiv:1512.08497 [hep-ph];
  N.~Bizot, S.~Davidson, M.~Frigerio and J.-L.~Kneur,
  arXiv:1512.08508 [hep-ph].

\bibitem{diphoton:dark matter}
  Y.~Mambrini, G.~Arcadi and A.~Djouadi,
  arXiv:1512.04913 [hep-ph];
  M.~Backovic, A.~Mariotti and D.~Redigolo,
  arXiv:1512.04917 [hep-ph];
  Y.~Bai, J.~Berger and R.~Lu,
  arXiv:1512.05779 [hep-ph];
  C.~Han, H.~M.~Lee, M.~Park and V.~Sanz,
  arXiv:1512.06376 [hep-ph];
  X.~J.~Bi, Q.~F.~Xiang, P.~F.~Yin and Z.~H.~Yu,
  arXiv:1512.06787 [hep-ph];
  M.~Bauer and M.~Neubert,
  arXiv:1512.06828 [hep-ph];
  D.~Barducci, A.~Goudelis, S.~Kulkarni and D.~Sengupta,
  arXiv:1512.06842 [hep-ph];
  U.~K.~Dey, S.~Mohanty and G.~Tomar,
  arXiv:1512.07212 [hep-ph];
  P.~S.~B.~Dev and D.~Teresi,
  arXiv:1512.07243 [hep-ph];
  H.~Han, S.~Wang and S.~Zheng,
  arXiv:1512.07992 [hep-ph];
  J.~C.~Park and S.~C.~Park,
  arXiv:1512.08117 [hep-ph].

\bibitem{diphoton:model independent}
  Q.~H.~Cao, Y.~Liu, K.~P.~Xie, B.~Yan and D.~M.~Zhang,
  arXiv:1512.05542 [hep-ph];
  J.~Chakrabortty, A.~Choudhury, P.~Ghosh, S.~Mondal and T.~Srivastava,
  arXiv:1512.05767 [hep-ph];
  J.~Bernon and C.~Smith,
  arXiv:1512.06113 [hep-ph];
  F.~P.~Huang, C.~S.~Li, Z.~L.~Liu and Y.~Wang,
  arXiv:1512.06732 [hep-ph];
  J.~S.~Kim, K.~Rolbiecki and R.~R.~de Austri,
  arXiv:1512.06797 [hep-ph];
  W.~S.~Cho, D.~Kim, K.~Kong, S.~H.~Lim, K.~T.~Matchev, J.~C.~Park and M.~Park,
  arXiv:1512.06824 [hep-ph];
  M.~Chala, M.~Duerr, F.~Kahlhoefer and K.~Schmidt-Hoberg,
  arXiv:1512.06833 [hep-ph];
  J.~Gao, H.~Zhang and H.~X.~Zhu,
  arXiv:1512.08478 [hep-ph].



\bibitem{Langacker:2008yv} 
 P.~Langacker,
 Rev.\ Mod.\ Phys.\  {\bf 81}, 1199 (2009)
 doi:10.1103/RevModPhys.81.1199
 [arXiv:0801.1345 [hep-ph]].

\bibitem{LandauYang} 
  L.~D.~Landau,
  Dokl.\ Akad.\ Nauk Ser.\ Fiz.\  {\bf 60}, no. 2, 207 (1948).
  doi:10.1016/B978-0-08-010586-4.50070-5;
  C.~N.~Yang,
  Phys.\ Rev.\  {\bf 77}, 242 (1950).
  doi:10.1103/PhysRev.77.242

\bibitem{Stueckelberg:1900zz} 
  E.~C.~G.~Stueckelberg,
  Helv.\ Phys.\ Acta {\bf 11}, 225 (1938).
  doi:10.5169/seals-110852

\bibitem{WeinbergII}
For a convenient reference, see S.~Weinberg, The quantum theory of fields II, Cambridge University Press (1996).

\bibitem{Georgi:1974sy} 
  H.~Georgi and S.~L.~Glashow,
  Phys.\ Rev.\ Lett.\  {\bf 32}, 438 (1974);
  for a review, see, e.g., 
  P.~Langacker,
  Phys.\ Rept.\  {\bf 72}, 185 (1981).

\bibitem{Giudice:2012zp} 
  G.~F.~Giudice, R.~Rattazzi and A.~Strumia,
  Phys.\ Lett.\ B {\bf 715}, 142 (2012)
  [arXiv:1204.5465 [hep-ph]];
  for related discussion, see, e.g.,
  M.~Ibe,
  JHEP {\bf 0908}, 086 (2009)
  [arXiv:0906.4667 [hep-ph]];
  N.~Haba, K.~Kaneta and R.~Takahashi,
  Phys.\ Lett.\ B {\bf 734}, 220 (2014)
  doi:10.1016/j.physletb.2014.05.016
  [arXiv:1309.1231 [hep-ph]];
  N.~Haba, K.~Kaneta and R.~Takahashi,
  Eur.\ Phys.\ J.\ C {\bf 74}, 2696 (2014)
  doi:10.1140/epjc/s10052-013-2696-z
  [arXiv:1309.3254 [hep-ph]];
  B.~Bajc and G.~Senjanovic,
  JHEP {\bf 0708}, 014 (2007)
  [hep-ph/0612029];
  B.~Bajc, M.~Nemevsek and G.~Senjanovic,
  Phys.\ Rev.\ D {\bf 76}, 055011 (2007)
  [hep-ph/0703080];
  A.~Arhrib, B.~Bajc, D.~K.~Ghosh, T.~Han, G.~Y.~Huang, I.~Puljak and G.~Senjanovic,
  Phys.\ Rev.\ D {\bf 82}, 053004 (2010)
  [arXiv:0904.2390 [hep-ph]];
  T.~S.~Ray, H.~de Sandes and C.~A.~Savoy,
  Phys.\ Lett.\ B {\bf 712}, 401 (2012)
  [arXiv:1112.6180 [hep-ph]];
  T.~Aizawa, M.~Ibe and K.~Kaneta,
  Phys.\ Rev.\ D {\bf 91}, no. 7, 075012 (2015)
  doi:10.1103/PhysRevD.91.075012
  [arXiv:1411.6044 [hep-ph]].


\bibitem{Bagger:1995bw}
  J.~Bagger, K.~T.~Matchev and D.~Pierce,
  Phys.\ Lett.\  B {\bf 348}, 443 (1995)
  [arXiv:hep-ph/9501277];
  D.~M.~Pierce, J.~A.~Bagger, K.~T.~Matchev and R.~j.~Zhang,
  Nucl.\ Phys.\  B {\bf 491}, 3 (1997)
  [arXiv:hep-ph/9606211].

\bibitem{Nishino:2012bnw} 
  H.~Nishino {\it et al.} [Super-Kamiokande Collaboration],
  Phys.\ Rev.\ D {\bf 85}, 112001 (2012)
  doi:10.1103/PhysRevD.85.112001
  [arXiv:1203.4030 [hep-ex]].

\bibitem{Giddings:1987cg} 
  S.~B.~Giddings and A.~Strominger,
  Nucl.\ Phys.\ B {\bf 306}, 890 (1988).
  doi:10.1016/0550-3213(88)90446-4;
  L.~F.~Abbott and M.~B.~Wise,
  Nucl.\ Phys.\ B {\bf 325}, 687 (1989).
  doi:10.1016/0550-3213(89)90503-8;
  S.~R.~Coleman and K.~M.~Lee,
  Nucl.\ Phys.\ B {\bf 329}, 387 (1990).
  doi:10.1016/0550-3213(90)90149-8


\bibitem{Basso:2010jm} 
  L.~Basso, S.~Moretti and G.~M.~Pruna,
  Phys.\ Rev.\ D {\bf 82}, 055018 (2010)
  doi:10.1103/PhysRevD.82.055018
  [arXiv:1004.3039 [hep-ph]].


\bibitem{Cirelli:2005uq} 
  M.~Cirelli, N.~Fornengo and A.~Strumia,
  Nucl.\ Phys.\ B {\bf 753}, 178 (2006)
  [hep-ph/0512090];
  M.~Cirelli, A.~Strumia and M.~Tamburini,
  Nucl.\ Phys.\ B {\bf 787}, 152 (2007)
  [arXiv:0706.4071 [hep-ph]];
  M.~Cirelli, R.~Franceschini and A.~Strumia,
  Nucl.\ Phys.\ B {\bf 800}, 204 (2008)
  [arXiv:0802.3378 [hep-ph]];
  M.~Cirelli and A.~Strumia,
  PoS IDM {\bf 2008}, 089 (2008)
  [arXiv:0808.3867 [astro-ph]];
  M.~Cirelli and A.~Strumia,
  New J.\ Phys.\  {\bf 11}, 105005 (2009)
  [arXiv:0903.3381 [hep-ph]].
\bibitem{Hisano:2006nn} 
  J.~Hisano, S.~Matsumoto, M.~Nagai, O.~Saito and M.~Senami,
  Phys.\ Lett.\ B {\bf 646}, 34 (2007)
  [hep-ph/0610249].

\bibitem{Harigaya:2014dwa} 
  K.~Harigaya, K.~Kaneta and S.~Matsumoto,
  Phys.\ Rev.\ D {\bf 89}, no. 11, 115021 (2014)
  doi:10.1103/PhysRevD.89.115021
  [arXiv:1403.0715 [hep-ph]].



\bibitem{Djouadi:2005gi} 
  A.~Djouadi,
  Phys.\ Rept.\  {\bf 457}, 1 (2008)
  doi:10.1016/j.physrep.2007.10.004
  [hep-ph/0503172];
  A.~Djouadi,
  Phys.\ Rept.\  {\bf 459}, 1 (2008)
  doi:10.1016/j.physrep.2007.10.005
  [hep-ph/0503173].

  
\bibitem{Martin:2009iq} 
  A.~D.~Martin, W.~J.~Stirling, R.~S.~Thorne and G.~Watt,
  Eur.\ Phys.\ J.\ C {\bf 63}, 189 (2009)
  doi:10.1140/epjc/s10052-009-1072-5
  [arXiv:0901.0002 [hep-ph]].

\bibitem{Ellis:2015oso} 
  J.~Ellis, S.~A.~R.~Ellis, J.~Quevillon, V.~Sanz and T.~You,
  arXiv:1512.05327 [hep-ph].



\bibitem{Khachatryan:2015wza} 
  V.~Khachatryan {\it et al.} [CMS Collaboration],
  JHEP {\bf 1506}, 116 (2015)
  doi:10.1007/JHEP06(2015)116
  [arXiv:1503.08037 [hep-ex]].


\bibitem{Carena:2004xs} 
  M.~Carena, A.~Daleo, B.~A.~Dobrescu and T.~M.~P.~Tait,
  Phys.\ Rev.\ D {\bf 70}, 093009 (2004)
  doi:10.1103/PhysRevD.70.093009
  [hep-ph/0408098];
  G.~Cacciapaglia, C.~Csaki, G.~Marandella and A.~Strumia,
  Phys.\ Rev.\ D {\bf 74}, 033011 (2006)
  doi:10.1103/PhysRevD.74.033011
  [hep-ph/0604111].

\bibitem{Chatrchyan:2012oaa} 
  S.~Chatrchyan {\it et al.} [CMS Collaboration],
  Phys.\ Lett.\ B {\bf 720}, 63 (2013)
  doi:10.1016/j.physletb.2013.02.003
  [arXiv:1212.6175 [hep-ex]].

\bibitem{Guo:2015lxa} 
  J.~Guo, Z.~Kang, P.~Ko and Y.~Orikasa,
  Phys.\ Rev.\ D {\bf 91}, no. 11, 115017 (2015)
  doi:10.1103/PhysRevD.91.115017
  [arXiv:1502.00508 [hep-ph]].
  

\bibitem{Godfrey:2013eta} 
  S.~Godfrey and T.~Martin,
  arXiv:1309.1688 [hep-ph].

\bibitem{Franceschini:2015kwy} 
  R.~Franceschini {\it et al.},
  arXiv:1512.04933 [hep-ph].
\bibitem{Craig:2015lra} 
  N.~Craig, P.~Draper, C.~Kilic and S.~Thomas,
  arXiv:1512.07733 [hep-ph].



\end{thebibliography}
\end{document}